\RequirePackage{fix-cm}
\documentclass[twocolumn,epjc3]{svjour3}
\smartqed  
\RequirePackage{graphicx}
\RequirePackage{color}

\RequirePackage{multirow}

\journalname{Eur. Phys. J. C}
\begin{document}

\title{Physics potential for the H$\rightarrow$ZZ$^{*}$ decay at the CEPC }



\author{  Ryuta Kiuchi\thanksref{addr1}
  \and
  Yanxi Gu\thanksref{addr2}  
  \and  
  Min Zhong\thanksref{addr2}
  \and
  Lingteng Kong\thanksref{addr3}
  \and
  Alex Schuy\thanksref{addr4}
  \and
  Shih-Chieh Hsu\thanksref{email2, addr4}
  \and 
  Xin Shi\thanksref{email1, addr1}
  \and
  Kaili Zhang\thanksref{addr1}
}


\thankstext{email1}{e-mail: shixin@ihep.ac.cn}
\thankstext{email2}{e-mail: schsu@uw.edu}


\institute{Institute of High Energy Physics, Chinese Academy of Science, Beijing 100049, China \label{addr1}
  \and   
  Department of Modern Physics, University of Science and Technology of China, Hefei 230026, China  \label{addr2}
  \and
  University of Chinese Academy of Sciences, Beijing, 100049, China \label{addr3}
  \and
  Department of Physics, University of Washington, Seattle 98195-1560, USA  \label{addr4}
}

\date{Received: date / Accepted: date}

\maketitle

\begin{abstract}
  The precision of the yield measurement of the Higgs boson decaying into a pair of $Z$ bosons process at the Circular
  Electron Positron Collider (CEPC) is evaluated. Including the recoil $Z$ boson associated with the Higgs production (Higgsstrahlung)
  total three $Z$ bosons are involved for this channel, from which final states characterized by the presence of a pair of leptons,
  quarks, and neutrinos are chosen for the signal.
  Two analysis approaches are compared and 
  the final precision of ${\sigma}_{\mathrm{ZH}}{\cdot}$BR($H \rightarrow ZZ^{*}$) is estimated to be 7.9\% using a multivariate analysis technique,
  based on boosted decision trees.
  The relative precision of the Higgs boson width, using this $H \rightarrow ZZ^{*}$ decay topology,
  is estimated by combining the obtained result with the precision of the inclusive \textit{ZH} cross section measurement.

\keywords{CEPC \and Higgs boson \and Higgs to ZZ}

\end{abstract}

\section{Introduction}
\label{section:introduction}

After the discovery of the Higgs boson~\cite{higgs_discovery_atlas,higgs_discovery_cms},
efforts are performed on measuring properties of the Higgs boson.
One of motivations of these studies is to obtain hints for physics beyond the Standard Model (SM),
whose existence is suggested by several experiment facts, such as dark matter, cosmological baryon-antibaryon asymmetry.
The Circular Electron Positron Collider (CEPC)~\cite{cepc_cdr1,cepc_cdr2} is a proposed future circular $e^{+}e^{-}$ collider,
with a main ring circumstance of $\sim$100 km. As a Higgs factory,
the CEPC is planned to operate at a center of mass energy $\sqrt{s}=240$ GeV with an integrated luminosity of $5.6\ {\rm ab^{-1}}$
corresponding to the production of more than $10^{6}$ Higgs bosons.
Hence it is expected to achieve an order of magnitude improvement on measurements of Higgs boson properties
as compared to the final LHC precision.

The Higgs production mechanisms in $e^{+}e^{-}$ collision at $\sqrt{s}=240$ GeV will be
the Higgsstrahlung process $e^{+}e^{-} \rightarrow Z^{*} \rightarrow ZH$ (hereafter, denoted as \textit{ZH} process)
and the vector boson fusion processes,
$e^{+}e^{-} \rightarrow W^{+*}W^{-*}{\nu}_{e}\bar{{\nu}}_{e} \rightarrow H{\nu}_{e}\bar{{\nu}}_{e}$
and
$e^{+}e^{-} \rightarrow Z^{*}Z^{*}e^{+}e^{-} \rightarrow He^{+}e^{-}$.
Among these processes, the \textit{ZH} process is predicted to have the largest cross section,
dominating over all of the others~\cite{cepc_white_paper}.
Therefore,  the \textit{ZH} production mode is going to provide series of the Higgs measurements, such as the inclusive \textit{ZH} process
cross section ${\sigma}_{\mathrm{ZH}}$,  using the recoil mass method against the $Z$ boson.
That $Z$ boson also serves as a tag of the \textit{ZH} process 
through reconstruction of objects decaying from the $Z$ boson.
Utilizing this tag information, the Higgs boson is clearly identified and thus
individual decay channels of the Higgs boson will be explored subsequently.    

The decay channel, where the Higgs boson decays into a pair of $Z$ bosons via the \textit{ZH} process, will be studied at the CEPC.
Like the other decay modes, the Branching ratio BR($H \rightarrow ZZ^{*}$) can be obtained from 
the measurement of the signal yield, since the yield allows to extract the observable ${\sigma}_{ZH}{\times}$BR($H \rightarrow ZZ^{*}$).
In addition, the Higgs boson width $\mathrm{\Gamma}_{H}$ can be inferred as well.
Under the assumption that the coupling structure follows the SM,
the branching ratio is proportional to 
BR$(H \rightarrow ZZ^{*}) = \mathrm{\Gamma}(H \rightarrow ZZ^{*})/\mathrm{\Gamma}_{H} \propto g^{2}_{HZZ}/\mathrm{\Gamma}_{H}$,
therefore, $\mathrm{\Gamma}_{H}$ can be deduced with
precision determined from the measurements of the coupling $g^{2}_{HZZ}$ (${\sigma}_{ZH} \propto g^{2}_{HZZ}$)
and the signal yield. Note that the vector boson fusion ${\nu}\bar{{\nu}}H$ process in combination with
measurements of the $H \rightarrow WW^{*}$ decay channel can also provide the $\mathrm{\Gamma}_{H}$ value independently,
hence the final value will be determined from the combination of the two measurements~\cite{cepc_white_paper}.

\begin{figure}[htb]
  \centering
  \includegraphics[width=0.6\hsize]{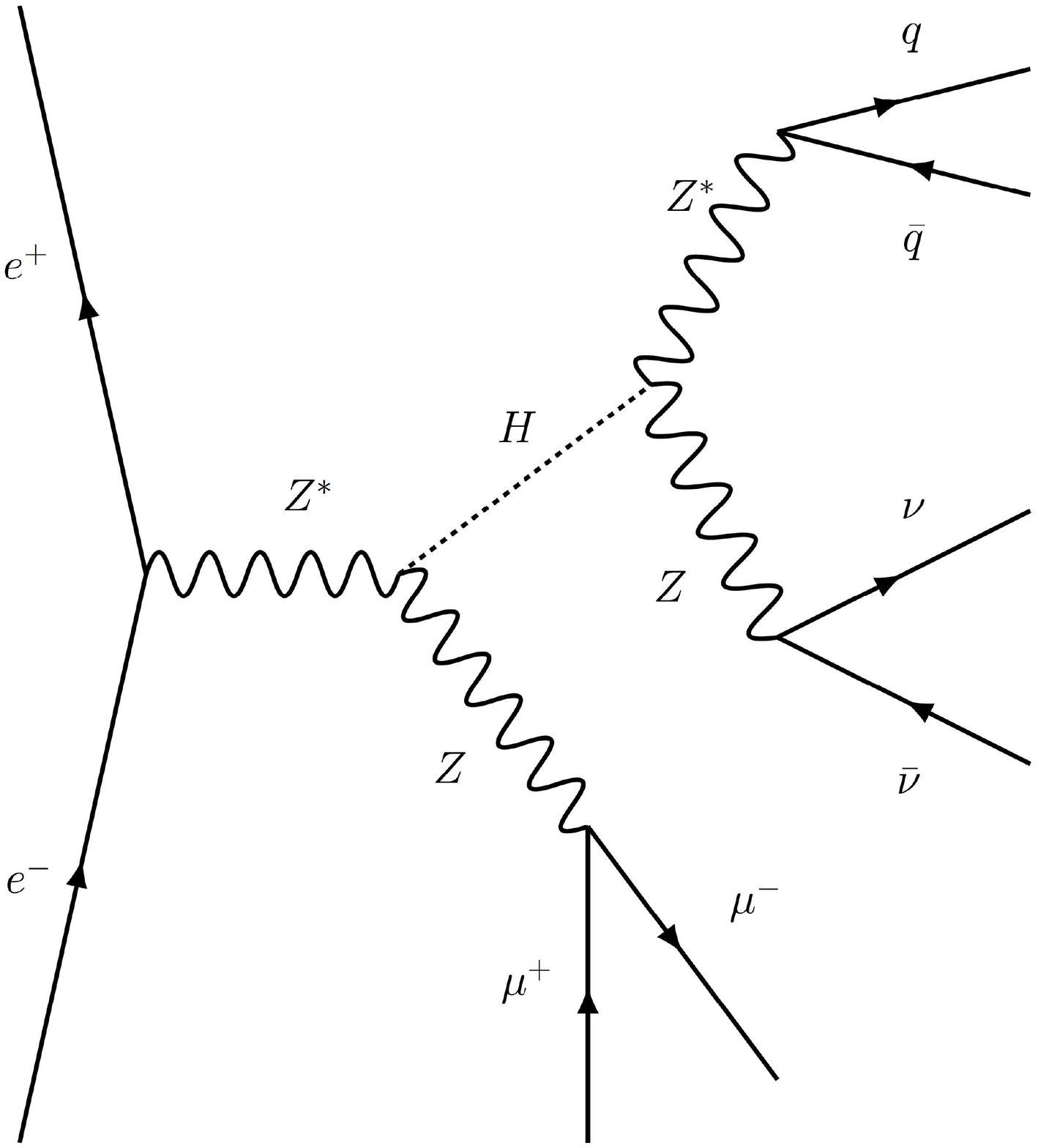}  
  \caption{
    Example Feynman diagram of the signal process which is characterized by the presence of
    a pair of muons, jets and neutrinos. In this example, the initial $Z$ boson recoiling against
    the Higgs boson is decaying into muons.
    Final states with all of cyclic permutation of the decay products from three $Z$ bosons are considered throughout this analysis.}    
    \label{figure:hzz_fyenman_diagram}
\end{figure}

The study of $H \rightarrow ZZ^{*}$ channel via the \textit{ZH} process has an unique feature among the other decays
that is originated from its event topology where two on-shell $Z$ bosons and one off-shell $Z$ boson are involved.
Considering that $Z$ bosons can decay to any fermion anti-fermion pair except a top quark pair, 
the topology diverges into lots of final states. The $H \rightarrow ZZ^{*} \rightarrow 4l$ decay is the
so-called ``golden channel'' of the Higgs boson study at the LHC, as it has the cleanest signature
of all the possible Higgs boson decay modes~\cite{higgs_to_4l_atlas,higgs_to_4l_cms}.
However, the statistics of this leptonic channel at the CEPC
may not allow to study the properties with required precision.
Conversely, fully hadronic channel can provide enough statistics, but difficulties in identifying and matching
jets with proper $Z$ bosons, as well as efficient separation from
the SM backgrounds have to be overcome.
Between these two extremes, the decay channels having a pair of leptons, two jets and two neutrinos
are most promising candidates for studying $H \rightarrow ZZ^{*}$ properties,
owing to its clear signature and larger branching fraction than the leptonic channel.
Therefore, this final state has been chosen as the signal for the evaluation of the $H \rightarrow ZZ^{*}$ properties.
Among charged leptons, muons have advantage on discrimination of 
isolated candidates from those produced via semi-leptonic decays of heavy flavor jets.
Therefore, the final states including a pair of muons are finally selected as the signal process: 
$Z \rightarrow {\mu}^{+}{\mu}^{-}$, $H {\rightarrow} ZZ^{*} \rightarrow {\nu}\bar{{\nu}}q\bar{q}$
(Fig.~\ref{figure:hzz_fyenman_diagram}) and its cyclic permutations,
$Z \rightarrow {\nu}\bar{{\nu}}$, $H \rightarrow ZZ^{*} \rightarrow q\bar{q}{\mu}^{+}{\mu}^{-}$ and 
$Z \rightarrow q\bar{q}$, $H \rightarrow ZZ^{*} \rightarrow {\mu}^{+}{\mu}^{-}{\nu}\bar{{\nu}}$,
where the $q$ represents all quark flavors except for the top quark.

In this article, we report on the estimation of relative precision of the yield measurement for
the $H \rightarrow ZZ^{*}$ decay at the CEPC using the signal processes characterized by
the presence of a pair of muons, jets and neutrinos.
In Sec.~\ref{section:detector_design}, we briefly introduce the CEPC detector design and the Monte Carlo
(MC) simulation scheme.
The details of the event selection on generated samples is described in Sec.~\ref{section:event_selection}.
The statistical procedure and results of the estimated precision of the signal yield is presented in Sec.~\ref{section:result}
followed by a brief discussion in Sec.~\ref{section:discussion}.
Finally, conclusions are summarized in Sec.~\ref{section:summary}. 

\section{Detector design and simulation samples}
\label{section:detector_design}

The CEPC will hosts two interaction points (IP) on the main ring,
where the detector at each IP records collision data under different
center of mass energies varying from $\sqrt{s}=91.2$ GeV as a Z factory to $\sqrt{s}=240$ GeV as a Higgs factory.
To fulfill the physics goals, a baseline concept of the detector is developed based on
the International Large Detector (ILD) concept~\cite{ilc_cdr} 
with further optimizations for the CEPC environment.
From the most inner sub-detector component, the detector concept is composed of a silicon vertex detector,
a silicon inner tracker consisting of micro strip detectors, a Time Projection Chamber (TPC), a silicon external tracker,
ultra-fine segmented calorimeters, an Electromagnetic CALorimeter (ECAL) and an Hadronic CALorimeter (HCAL),
a 3T superconducting solenoid, and a muon detector~\cite{cepc_cdr2}.

The CEPC simulation software package implements the baseline concept detector geometry.
Events for the SM processes are generated by the Whizard~\cite{whizard} including the Higgs boson signal,
where the detector configuration and response is handled by the GEANT4-based simulation framework, MokkaPlus~\cite{mokka}.
Modules for digitization of the signals at each sub detector creates the hit information.
Particle reconstruction has been taken place with the Arbor algorithm, which builds the
reconstructed particles using calorimeter and track information~\cite{arbor}.
A set of MC samples at $\sqrt{s}=240$ GeV has been generated with this scheme where the Higgs boson signal
also contain the $WW/ZZ$ fusion processes.
All of the SM background samples, which can be classified according to number of fermions in their final states,
two-fermion processes ($e^{+}e^{-} \rightarrow f\bar{f}$)
and four-fermion processes ($e^{+}e^{-} \rightarrow f\bar{f}f\bar{f}$), are produced as well.
More details about the samples and their classification can be found in Ref.~\cite{cepc_sample_classification}.

\section{Event selection}
\label{section:event_selection}

Event selection is performed in several stages.
The pre-selection builds higher-level objects, such as isolated muons, jets, and missing four momentum
from the Particle Flow (PF) objects which are reconstructed by the ArborPFA.
The isolation requirements on muons, identified by the PFs, are imposed.
For muons with energy higher than 3 GeV,  tracks inside of a cone with a half-angle ${\theta}$
around the candidate are examined and it is identified as an isolated muon, when a ratio of
the energy of the muon candidate to summation of the energy
from all of the tracks except for the candidate in a volume defined by the cone
is greater than 10 with ${\cos {\theta}}=0.98$.
Jets are clustered from the PFs but except for isolated lepton candidates,
using the $k_{t}$ algorithm for the $e^{+}e^{-}$ collision ($ee-kt$) with the FastJet package~\cite{fastjet}.
Exclusive requirement ($N_{jet}=2$) on number of jets is imposed.
Events are requested to have a pair of isolated muons of positive and negative charged, 
and two jets successfully clustered.

The events satisfying the pre-selection criteria are separated into six categories.
Depending on which physics objects ($\mu\mu$/$q\bar{q}$/$\nu\bar{\nu}$) form the tagged $Z$ boson (hereafter denoted it as initial $Z$ boson) ,
the signal samples can be classified into three categories.
Furthermore,
distinguishing the status between having a pair of objects
suppose to be decaying from the on-shell $Z$ boson and from the off-shell $Z$ boson where $H \rightarrow ZZ^{*}$
decay is assumed, 
enhances the efficiency of the event selection
by applying different selection criteria for each respectively.
Following notation is adopted for denoting each category: ${\mu}{\mu}{\mathrm{H}}{\nu}{\nu}qq$ (${\mu}{\mu}{\mathrm{H}}qq{\nu}{\nu}$)
category is defined
to be most sensitive to signal events
having reconstructed invariant mass $M_{{\mu}{\mu}}$ of two muons in the range 80-100 GeV 
where two top characters in the notation represent a pair of muons decaying from the initial $Z$ boson,
with the reconstructed invariant mass of missing term $M_{\mathrm{miss}}$ due to escaping neutrinos
is larger (smaller) than dijet invariant mass $M_{jj}$.
The mass range of the initial $Z$ boson for the other categories are chosen as 75-110 GeV for
${\nu}{\nu}{\mathrm{H}}{\mu}{\mu}qq$ and ${\nu}{\nu}{\mathrm{H}}qq{\mu}{\mu}$ categories,
75-105 GeV for $qq{\mathrm{H}}{\nu}{\nu}{\mu}{\mu}$ and $qq$H${\mu}{\mu}{\nu}{\nu}$ categories,
taking into account the reconstructed mass resolution for this analysis.
The recoil mass against the initial $Z$ boson is required to be in the range of 110-140 GeV.
To ensure that the events are separated into categories exclusively, further requirements on recoil mass distributions
of a pair of objects are applied that is described later.

On total six categories,
${\mu}{\mu}{\mathrm{H}}{\nu}{\nu}qq$, ${\mu}{\mu}{\mathrm{H}}qq{\nu}{\nu}$,
${\nu}{\nu}{\mathrm{H}}{\mu}{\mu}qq$, ${\nu}{\nu}{\mathrm{H}}qq{\mu}{\mu}$,
$qq{\mathrm{H}}{\nu}{\nu}{\mu}{\mu}$, $qq{\mathrm{H}}{\mu}{\mu}{\nu}{\nu}$,
further event selection criteria are optimized separately.
Two different analysis approaches are exploited for this stage, the one where requirements are imposed on
a set of kinematic variables (referred to ``cut-based'' analysis) and the one which uses a multivariate analysis technique,
based on the boosted decision tree (BDT) implemented within scikit-learn package~\cite{scikit-learn},
in order to achieve better separation between signal and background (referred to ``BDT'' analysis).

For the cut-based analysis,
the signal to background ratio is maximized by the following requirements.
The invariant mass $M_{{\mu}{\mu}}$ of the two muons, the invariant mass $M_{jj}$ of two jets 
and the missing mass $M_{\mathrm{miss}}$ are required to fall into the mass window around the $Z$ ($Z^{*}$) boson.
Number of particle flow objects $N_{\mathrm{PFO}}$  in the event is required to be larger than a threshold value,
which is decided by the condition whether jets are originated from an on-shell $Z$ boson or not,
as well as suppression of background contributions where the jets are reconstructed from any objects other than quark seeds. 
Cut on the polar angle of the sum of all visible particles ${\cos {\theta}_{\mathrm{vis}}}$ is applied to further reject background processes,
such as two-fermion components which tends to be back-to-back along the beam axis.
The angle between the di-muons and di-jets systems 
${\Delta}{\phi}_{ZZ}$ is used to reduce background components as well.
Kinematic properties of two on-shell $Z$ bosons has significant overlap at $\sqrt{s}=240$ GeV.
As a result, a signal process and its \textit{conjugate} process that is the signal process by exchanging decay objects from on-shell $Z$ bosons,
e.g.  
$Z \rightarrow {\mu}^{+}{\mu}^{-}$, $H \rightarrow (Z{\rightarrow}q\bar{q}, Z^{*}{\rightarrow}{\nu}\bar{{\nu}})$
and
$Z \rightarrow q\bar{q}$, $H \rightarrow (Z{\rightarrow}{\mu}^{+}{\mu}^{-}, Z^{*}{\rightarrow}{\nu}\bar{{\nu}})$,
have considerable overlaps in the kinematical phase space.
To ensure that the events are grouped into mutually exclusive categories which are optimized based on their kinematic properties,
two exclusive regions in the two-dimensional phase space of recoil mass distributions of di-objects,  are defined and are used to further restrict categories.
For example, in the $M^{\mathrm{recoil}}_{{\mu}{\mu}}$-$M^{\mathrm{recoil}}_{jj}$ phase space,
a region covering majority of 
$Z \rightarrow {q}\bar{q}$, $H \rightarrow (Z{\rightarrow}{\mu}^{+}{\mu}^{-}, Z^{*}{\rightarrow}{\nu}\bar{{\nu}})$
signal events is defined as 

\begin{eqnarray*}
\begin{array}{llrr}
M^{\mathrm{recoil}}_{{\mu}{\mu}}-M_{H} & > &   \left| M^{\mathrm{recoil}}_{jj}-M_{H} \right|     &   \quad (M^{\mathrm{recoil}}_{{\mu}{\mu}} > M_{H})   \\
                                                                           & < &  - \left| M^{\mathrm{recoil}}_{jj}-M_{H} \right|    &   \quad (M^{\mathrm{recoil}}_{{\mu}{\mu}} < M_{H})  
\end{array}
\end{eqnarray*}
where $M_{H}$ represents the Higgs boson mass of 125 GeV.
A requirement, denoted by ``\textit{not-qqHZZ}'', has been added to the cut sequence for ${\mu}{\mu}{\mathrm{H}}qq{\nu}{\nu}$ category
where events are rejected if a set of reconstructed recoil mass ($M^{\mathrm{recoil}}_{{\mu}{\mu}}$, $M^{\mathrm{recoil}}_{jj}$) satisfies above condition.
Similarly, total two kinds of ``\textit{not-}xx\textit{HZZ}'' (xx$:{\mu}{\mu}$ or ${\nu}{\nu}$ or $qq$) cuts are added in the selection for each category.
Table~\ref{table:cut_criteria} summaries the selection criteria applied across all the categories considered.

\begin{table*}[hbtp]
  \caption{Overview of the requirements applied when selecting events (cut-based). }
  \label{table:cut_criteria}
  \centering
  \begin{tabular}{lcccccc}
    \hline
    \multicolumn{7}{c}{Pre-selections} \\
    \hline
    \multicolumn{7}{l}{$N(l)=2$, where leptons(l) should pass the isolation criteria} \\
    \multicolumn{7}{l}{$N({\mu}^{+})=1$, $N({\mu}^{-})=1$ with $E({\mu}^{\pm})>3$ GeV } \\
    \multicolumn{7}{l}{$N(jet)=2$} \\    
    \hline
    \multicolumn{1}{l}{  Selection (Cut-based)  }  &    
    \multicolumn{1}{c}{  ${\mu}{\mu}{\mathrm{H}}{\nu}{\nu}qq$   }  &
    \multicolumn{1}{c}{  ${\mu}{\mu}{\mathrm{H}}qq{\nu}{\nu}$   }  &        
    \multicolumn{1}{c}{  ${\nu}{\nu}{\mathrm{H}}{\mu}{\mu}qq$   }  &
    \multicolumn{1}{c}{  ${\nu}{\nu}{\mathrm{H}}qq{\mu}{\mu}$  }  &
    \multicolumn{1}{c}{  $qq{\mathrm{H}}{\nu}{\nu}{\mu}{\mu}$  }  &
    \multicolumn{1}{c}{  $qq{\mathrm{H}}{\mu}{\mu}{\nu}{\nu}$  }  \\
    \hline
    Mass order                                                          &
    $M_{\mathrm{miss}} > M_{jj}$ & $M_{\mathrm{miss}} < M_{jj}$  &
    $M_{{\mu}{\mu}} > M_{jj}$     & $M_{{\mu}{\mu}} < M_{jj}$      &
    $M_{\mathrm{miss}} > M_{{\mu}{\mu}}$ & $M_{\mathrm{miss}} < M_{{\mu}{\mu}}$ \\
    
    $M_{{\mu}{\mu}}$ (GeV)                                      &  \multicolumn{2}{c}{ [80, 100] }    &  [60, 100]      & [10, 60]     & [15, 55]      & [75, 100]       \\
    $M_{jj}$ (GeV)                                                   &  [15, 60]       &   [60, 105]    &  [10, 55]       & [60, 100]   & \multicolumn{2}{c}{ [75, 105] }      \\
    $M_{\mathrm{miss}}$ (GeV)                                 &  [75, 105]     &   [10, 55]      &  \multicolumn{2}{c}{ [75, 110] } & [70, 110]    & [10, 50]          \\
    $M^{\mathrm{recoil}}_{{\mu}{\mu}}$ (GeV)           &  \multicolumn{2}{c}{ [110, 140] }   &   -                &  -                & [175, 215]  & [115, 155]       \\
    $M_{\mathrm{vis}}$ (GeV)                                   &   -                  &   [175, 215]   &   \multicolumn{2}{c}{ [110, 140] } & [115, 155]  & [185, 215]     \\
    $M^{\mathrm{recoil}}_{jj}$ (GeV)                         &  [185, 220]    &  -                   &   -                  &  -                & \multicolumn{2}{c}{ [110, 140] }    \\
    $N_{\mathrm{PFO}}$                                             &  [20, 90]       &   [30, 100]     &   [20, 60]       &  [30, 100]   & [40, 95]  &  [40, 95]       \\
    $\left| {\cos {\theta}_{\mathrm{vis}}} \right|$    &  \multicolumn{6}{c}{$<0.95$}                                                                                        \\        
    ${\Delta}{\phi}_{ZZ}$  (degree)                      &  [60, 170]     &   [60, 170]     &   $<135$       &   $<135$    &  -             &  [120, 170]            \\

    Region masking                                                &
    \multicolumn{2}{c}{\textit{not-$\nu\nu$HZZ} \& \textit{not-qqHZZ}}  &
    \multicolumn{2}{c}{\textit{not-$\mu\mu$HZZ} \& \textit{not-qqHZZ}} &
    \multicolumn{2}{c}{\textit{not-$\nu\nu$HZZ} \& \textit{not-$\mu\mu$HZZ}}                   \\
    
    \hline
    
  \end{tabular}
\end{table*}

The signal and background reduction efficiencies together with expected number of events running at
$\sqrt{s}=240$ GeV corresponding to a total integrated luminosity of 5.6 ab$^{-1}$ after the event selection
are listed in the Table~\ref{table:analysis_eff}.
For the signal events,
Table~\ref{table:analysis_eff} reports the number of events for the dominant and sub-dominant signal process separately,
where the sub-dominant signal process in the category is always the \textit{conjugate} process. 
The rest of signal processes other than the two processes are not listed in the table since their contributions are found to be very small.
In general, the analysis achieves a strong background rejection,
while the signal selection efficiencies of approximately 30\% and higher are kept.
The major background which are common in all categories is the other Higgs decays.
Four-fermion processes, particularly $e^{+}e^{-} \rightarrow ZZ \rightarrow {\mu}^{+}{\mu}^{-}q\bar{q}$ component in
both of the ${\mu}{\mu}$H$qq{\nu}{\nu}$ and $qq$H${\mu}{\mu}{\nu}{\nu}$ categories,
and $e^{+}e^{-} \rightarrow ZZ \rightarrow {\tau}^{+}{\tau}^{-}q\bar{q}$ component in both of the
${\nu}{\nu}{\mathrm{H}}qq{\mu}{\mu}$ and $qq{\mathrm{H}}{\nu}{\nu}{\mu}{\mu}$ categories,
have large contributions due to similarity of their kinematics.

\begin{table*}[hbtp]
  \caption{Summary of the selection efficiency $\epsilon$ and the number of expected events N$_{evt.}$ for each category
    after the final event selection in the cut-based analysis.. }
  \label{table:analysis_eff}
  \centering
  \begin{tabular}{l|rr|rr|rr}
    \hline
    \multicolumn{1}{c}{  (Cut-based)  }  &
    \multicolumn{2}{c}{  ${\mu}{\mu}{\mathrm{H}}{\nu}{\nu}qq$   }  &
    \multicolumn{2}{c}{  ${\mu}{\mu}{\mathrm{H}}qq{\nu}{\nu}$   }  &
    \multicolumn{2}{c}{  ${\nu}{\nu}{\mathrm{H}}{\mu}{\mu}qq$   }  \\
    \hline
    \multicolumn{1}{c}{Process}             &  \multicolumn{1}{c}{$\epsilon$ [\%]} & \multicolumn{1}{r}{N$_{evt.}$} &
    \multicolumn{1}{c}{$\epsilon$ [\%]} & \multicolumn{1}{r}{N$_{evt.}$} &  \multicolumn{1}{c}{$\epsilon$ [\%]} & \multicolumn{1}{r}{N$_{evt.}$} \\
    \hline    
    Signal (``dominant'')         & 38                                   &  53    & 36                                  & 50     &   54                                   &    76      \\
    Signal (``sub'')           & 6                                      &   8    & 10                                  & 14     &     6                                    &      9      \\    
    Higgs decays Bg.       &  $2.2{\cdot}10^{-3}$    & 25    & $7.0{\cdot}10^{-2}$     & 794   &   $5.3{\cdot}10^{-4}$       &     6      \\
    SM four-fermion Bg.  &  $3.7{\cdot}10^{-6}$    &    4   & $4.9{\cdot}10^{-4}$     &  520   &  $5.6{\cdot}10^{-6}$       &      6      \\
    SM two-fermion Bg.  &  0 &    0  & 0   &  0    &  0   &      0       \\    
    \hline

    \multicolumn{1}{c}{    }  &
    \multicolumn{2}{c}{  ${\nu}{\nu}{\mathrm{H}}qq{\mu}{\mu}$  }  &    
    \multicolumn{2}{c}{  $qq{\mathrm{H}}{\nu}{\nu}{\mu}{\mu}$  }  &
    \multicolumn{2}{c}{  $qq{\mathrm{H}}{\mu}{\mu}{\nu}{\nu}$  }  \\
    \hline
    \multicolumn{1}{c}{Process}             &  \multicolumn{1}{c}{$\epsilon$ [\%]} & \multicolumn{1}{r}{N$_{evt.}$} &
    \multicolumn{1}{c}{$\epsilon$ [\%]} & \multicolumn{1}{r}{N$_{evt.}$} &  \multicolumn{1}{c}{$\epsilon$ [\%]} & \multicolumn{1}{r}{N$_{evt.}$} \\
    \hline        
    Signal (``dominant'')          & 36                                   &   51   & 26                                    & 37      &  23                                  &    32   \\
    Signal (``sub'')             &  8                                    &   11   &  7                                    &  10      &   4                                  &      6   \\    
    Higgs decays Bg.       &  $1.0{\cdot}10^{-2}$     & 114   &  $2.4{\cdot}10^{-2}$     & 275     &  $1.4{\cdot}10^{-2}$     &  160  \\
    SM four-fermion Bg.   &  $4.3{\cdot}10^{-5}$     &   46   &  $1.5{\cdot}10^{-4}$     & 157    &  $1.8{\cdot}10^{-4}$     &  190  \\
    SM two-fermion Bg.   & 0     &    0   & 0     & 0      &  0  &       0  \\     
    \hline
    
  \end{tabular}
\end{table*}

For the BDT analysis, simpler selection criteria are applied prior to the BDT discrimination.
The invariant and recoil mass of the initial $Z$ boson which is reconstructed from di-objects
are required to be in the region of the signal mass window. 
The selection requirements on the number of particle flow objects and the polar angle of the sum of all visible particles
are also applied as used in the cut-based analysis. 

\begin{figure*}[hbtp]
  \centering

    \begin{tabular}{cc}
      
      \begin{minipage}{0.40\hsize}
        \centering
          \includegraphics[width=\hsize]{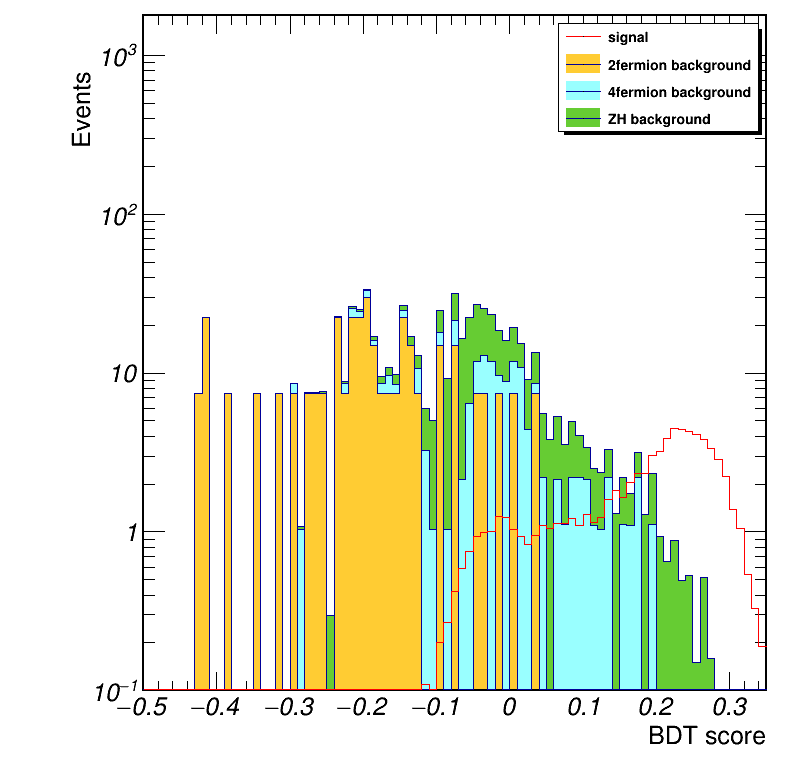}
      \end{minipage}
      &
      \begin{minipage}{0.40\hsize}
        \centering
          \includegraphics[width=\hsize]{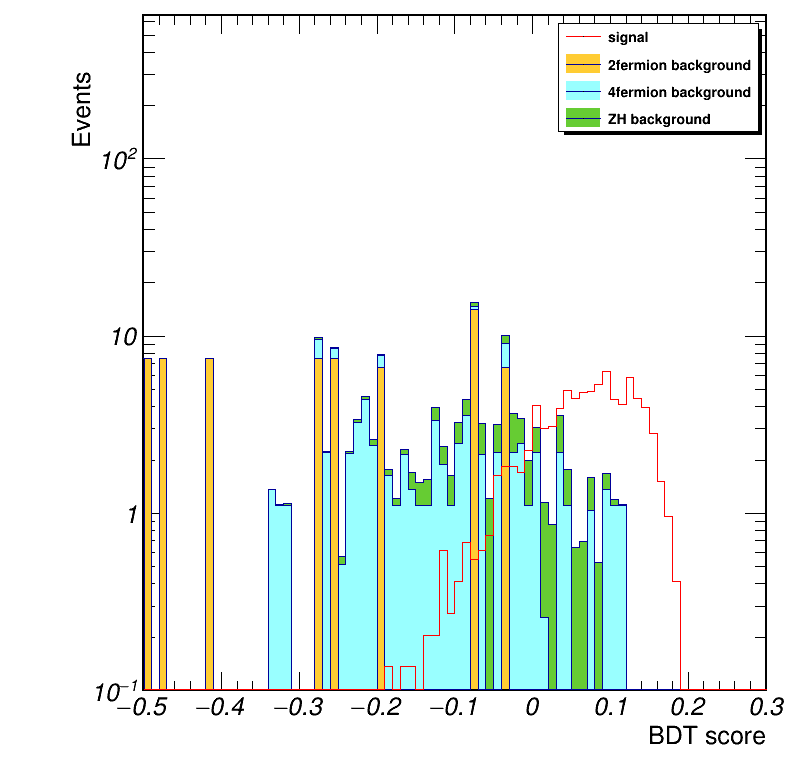}          
      \end{minipage}      
      
    \end{tabular}

    \caption{(color online) BDT score distributions for two of most sensitive categories:
      ${\mu}{\mu}{\mathrm{H}}{\nu}{\nu}qq^{\mathrm{mva}}$ (left) and
      ${\nu}{\nu}{\mathrm{H}}{\mu}{\mu}qq^{\mathrm{mva}}$ (right).
      The signal distribution is shown with a red histogram while background contributions,
      ZH (green), four-fermion (cyan) and two-fermion (yellow), are drawn.
    }
    \label{figure:bdt_score}
    
\end{figure*}

A boosted decision tree is then trained on remaining signal and background events for each category separately.
The boosting algorithm utilized in this analysis is the AdaBoost scheme~\cite{ada_boost}.
The input variables to the BDT are defined as follows:
\begin{itemize}
\item[\textbullet]  $M_{{\mu}{\mu}}$, $M_{jj}$, $M_{\mathrm{miss}}$ : invariant mass of di-objects
\item[\textbullet]  $N_{\mathrm{PFO}}$ : number of PFOs
\item[\textbullet]  ${\cos {\theta}_{\mathrm{vis}}}$: polar angle of the sum of all visible particles
\item[\textbullet]  ${\Delta}{\phi}_{ZZ}$ : angle between a $Z$ boson reconstructed from the two muons and that reconstructed from the two jets
\item[\textbullet]  $M^{\mathrm{recoil}}_{jj}$, $M_{\mathrm{vis}}$ : recoil mass of the di-jets and invariant mass of all visible particles (for ${\mu}{\mu}{\mathrm{H}}{\nu}{\nu}qq$ and ${\mu}{\mu}{\mathrm{H}}qq{\nu}{\nu}$ categories)
\item[\textbullet]  $M^{\mathrm{recoil}}_{jj}$, $M^{\mathrm{recoil}}_{{\mu}{\mu}}$ : recoil mass of the di-jets and the di-muons (for ${\nu}{\nu}{\mathrm{H}}{\mu}{\mu}qq$ and ${\nu}{\nu}{\mathrm{H}}qq{\mu}{\mu}$ categories)
\item[\textbullet]  $M^{\mathrm{recoil}}_{{\mu}{\mu}}$, $M_{\mathrm{vis}}$ :  recoil mass of the di-muons and invariant mass of all visible particles (for $qq{\mathrm{H}}{\nu}{\nu}{\mu}{\mu}$ and $qq{\mathrm{H}}{\mu}{\mu}{\nu}{\nu}$ categories)
\item[\textbullet]  $P_{\mathrm{vis}}$,   $P_{t, \mathrm{vis}}$ : magnitude of the momentum and transverse momentum from summation of all visible particles
\item[\textbullet]  $E^{leading}_{j}$, $E^{sub.}_{j}$ : energy of the leading jet and the sub-leading jet
\item[\textbullet]  $P^{leading}_{t, j}$, $P^{sub.}_{t, j}$ : magnitude of transverse momentum of the leading jet and the sub-leading jet 
\end{itemize}
The BDT analysis exploits the increased sensitivity by combining these 14 input variables
into the final BDT discriminant.
Fig.~\ref{figure:bdt_score} shows the obtained BDT score distributions for signal and background samples.
For the final separation of signal and background events, the cut value on the BDT score is chosen so as to maximize a significance measure $S/\sqrt{S+B}$,
where for a chosen cut, $S$ ($B$) is the number of signal (background) events above this cut.
The cut values as well as the other selection criteria are summarized in Table~\ref{table:cut_criteria_bdt-based}.

\begin{table*}[hbtp]
  \caption{Overview of the requirements applied when selecting events (BDT-based). }
  \label{table:cut_criteria_bdt-based}
  \centering
  \begin{tabular}{lcccccc}
    \hline
    \multicolumn{7}{c}{Pre-selections} \\
    \hline
    \multicolumn{7}{l}{$N(l)=2$, where leptons(l) should pass the isolation criteria} \\
    \multicolumn{7}{l}{$N({\mu}^{+})=1$, $N({\mu}^{-})=1$ with $E({\mu}^{\pm})>3$ GeV } \\
    \multicolumn{7}{l}{$N(jet)=2$} \\
    \hline
    \multicolumn{1}{l}{  Selection (MVA)}  &
    \multicolumn{1}{c}{  ${\mu}{\mu}{\mathrm{H}}{\nu}{\nu}qq$   }  &
    \multicolumn{1}{c}{  ${\mu}{\mu}{\mathrm{H}}qq{\nu}{\nu}$   }  &
    \multicolumn{1}{c}{  ${\nu}{\nu}{\mathrm{H}}{\mu}{\mu}qq$   }  &
    \multicolumn{1}{c}{  ${\nu}{\nu}{\mathrm{H}}qq{\mu}{\mu}$  }   &
    \multicolumn{1}{c}{  $qq{\mathrm{H}}{\nu}{\nu}{\mu}{\mu}$  }   &
    \multicolumn{1}{c}{  $qq{\mathrm{H}}{\mu}{\mu}{\nu}{\nu}$  }   \\
    \hline
    Mass order                                                          &
    $M_{\mathrm{miss}} > M_{jj}$ & $M_{\mathrm{miss}} < M_{jj}$  &
    $M_{{\mu}{\mu}} > M_{jj}$      & $M_{{\mu}{\mu}} < M_{jj}$      &
    $M_{\mathrm{miss}} > M_{{\mu}{\mu}}$ & $M_{\mathrm{miss}} < M_{{\mu}{\mu}}$ \\    

    $M_{{\mu}{\mu}}$ (GeV)                    &  \multicolumn{2}{c}{ [80,100] }  &  -            &  -             & -            & -               \\    
    $M_{jj}$ (GeV)                                  &  -            &  -             &  -            &  -             & \multicolumn{2}{c}{ [75, 105] }    \\
    $M_{\mathrm{miss}}$ (GeV)                &  -            &  -             &  \multicolumn{2}{c}{ [75, 110] } & -            & -               \\
    $M^{\mathrm{recoil}}_{{\mu}{\mu}}$ (GeV)  &  \multicolumn{2}{c}{ [110, 140] }   &  -            &  -             & -            & -               \\
    $M_{\mathrm{vis}}$ (GeV)                 &  -            &  -             &  \multicolumn{2}{c}{ [110, 140] }    & -            & -               \\
    $M^{\mathrm{recoil}}_{jj}$ (GeV)      &  -            &  -             &  -            &  -             & \multicolumn{2}{c}{ [110, 140] }     \\
    $N_{\mathrm{PFO}}$                        &  [20, 90]     &  [30, 100]     &  [20, 60]     &  [30, 100]     & [40, 95]     & [40, 95]       \\
    $\left| {\cos {\theta}_{\mathrm{vis}}} \right|$    &  \multicolumn{6}{c}{$<0.95$}                                                           \\
    Region masking                                                &
    \multicolumn{2}{c}{\textit{not-$\nu\nu$HZZ} \& \textit{not-qqHZZ}}  &
    \multicolumn{2}{c}{\textit{not-$\mu\mu$HZZ} \& \textit{not-qqHZZ}} &
    \multicolumn{2}{c}{\textit{not-$\nu\nu$HZZ} \& \textit{not-$\mu\mu$HZZ}}                   \\
    
    $BDT \ score$                             &  $>0.14$      &  $>0.01$       &  $>-0.01$     &  $>-0.01$       & $>-0.04$     & $>-0.01$        \\    
    \hline
  \end{tabular}
\end{table*}

\section{Result}
\label{section:result}

\begin{figure*}[hbtp]
  \centering

    \begin{tabular}{cc}
      
      \begin{minipage}{0.40\hsize}
        \centering
          \includegraphics[width=\hsize]{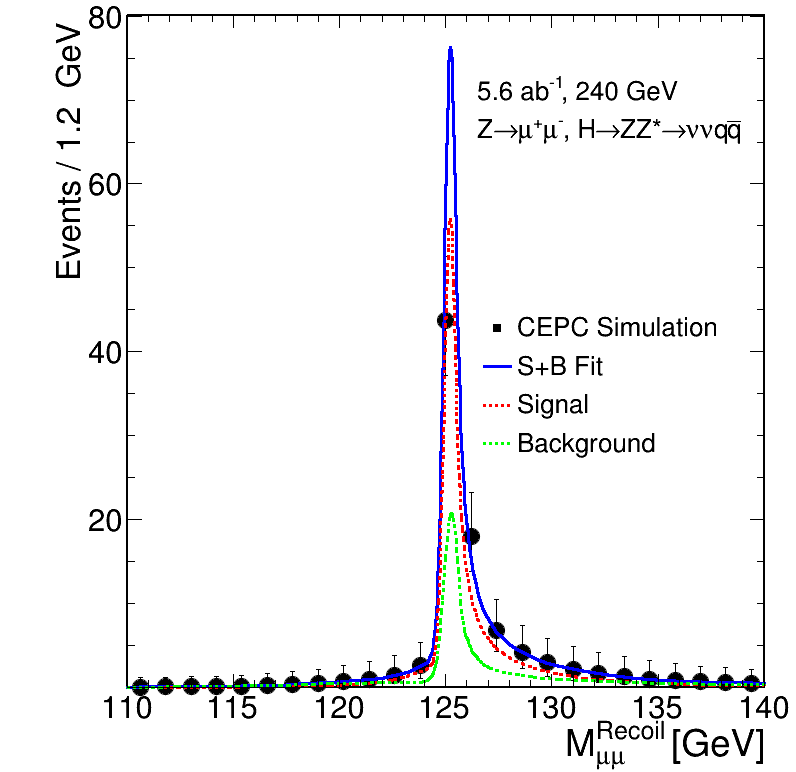}
      \end{minipage}
      &    
      \begin{minipage}{0.40\hsize}
        \centering        
          \includegraphics[width=\hsize]{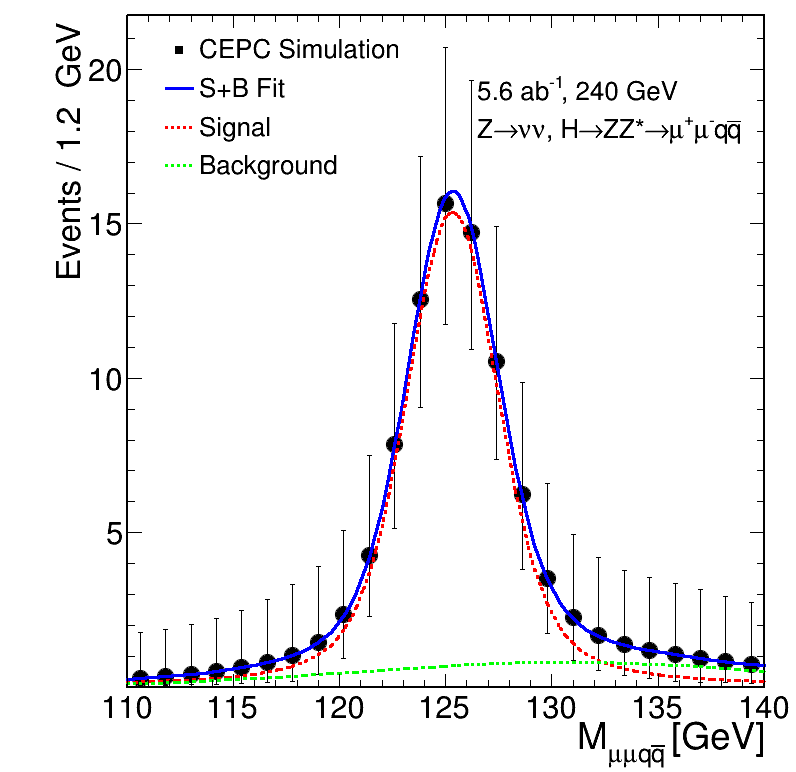}          
      \end{minipage}      
      
    \end{tabular}

    \caption{(color online) Recoil mass distributions in ${\mu}{\mu}{\mathrm{H}}{\nu}{\nu}qq^{\mathrm{cut}}$ (left) and
      ${\nu}{\nu}{\mathrm{H}}{\mu}{\mu}qq^{\mathrm{cut}}$ (right) categories in the cut-based analysis.
      The black dots represent the predicted results at the CEPC and the solid blue
      line shows the fitted model which is broken down into signal (dashed red line) and background (dashed green line) components.}
    \label{figure:likelihood_fitting}
    
\end{figure*}

An unbinned maximum likelihood fit is performed to extract the signal yield for each of six categories.
The obtained signal and background distributions of recoil mass spectrum $M_{Z}^{recoil}$ against the initial $Z$ boson
in the range 110-140 GeV, are added to make up a pseudo-experimental result,
while the likelihood template is constructed from sum of the Probability Density Function (PDF) describing the distributions of $M_{Z}^{recoil}$
for the signal and the background individually.
The normalized distribution of $M_{Z}^{recoil}$ for signal events in a category is described by sum of a double sided Crystal Ball function
and small Gaussian tails for the signal process with the initial $Z$ boson decaying to di-muon and 
a Breit-Wigner function convolved with a Gaussian for the rest of signal processes.
For the SM background components, a continuous PDF is constructed using the kernel density estimation technique~\cite{RooKeysPdf_ref}
for each component.
The background events from the other Higgs decay channels are modeled by the same PDF as the signal in terms of decay objects from the initial $Z$ boson,
except for the channels having small number of events ($<20$) where a PDF from the kernel density estimation is used
to describe the shape.The background components mentioned above are combined according to their fraction and are normalized to
the number of events left in the category.
The template model used to the likelihood fit is then expressed as ${\mu}{\cdot}N_{sig}{\cdot}f_{sig}+N_{bkg}{\cdot}f_{bkg}$,
where $f_{sig}$ $ (f_{bkg})$, $N_{sig}$ $ (N_{bkg})$ are the combined PDF and total number of events for signal (background) events,
$\mu$ is a free parameter determined by the fit.
Note that nuisance parameters, such as uncertainty of the total luminosity, are fixed to the expected values.
The recoil mass distribution together with the fitting results for two of the most sensitive categories
is shown in Fig.~\ref{figure:likelihood_fitting}.

The number of expected signal events can be simply represented by
$N_{sig} = \mathcal{L}{\cdot}{\epsilon}{\cdot}{\sigma}_{ZH}{\cdot}\mathrm{BR}(H \rightarrow ZZ^{*}) \cdot \prod_{X={\mu},{\nu},q} \mathrm{BR}(Z \rightarrow X\bar{X})$,
where $\mathcal{L}$ is the total luminosity and $\epsilon$ represents efficiencies including
the detector acceptance and the analysis selection.
The uncertainty of the fitting parameter $\mu$ is then regarded as the uncertainty of ${\sigma}_{ZH}{\cdot}$BR($H \rightarrow ZZ^{*}$)
by neglecting other systematic uncertainties.
Table~\ref{table:final_result} summarizes the derived relative precision on the product of the inclusive $ZH$ cross section
and the branching ratio ${\Delta}({\sigma}{\cdot}$BR)/(${\sigma}{\cdot}$BR) from the cut-based analysis
and the BDT analysis. The bottom row shows the combined precision that is calculated from the standard error
of the weighted mean, ${\sigma}=1/\sqrt{\sum^{n}_{i=1}{\sigma}^{-2}_{i}}$, where  ${\sigma}_{i}$
is the precision for each category.
The final result for the relative statistical uncertainty of the 
${\sigma}_{ZH}$ ${\times}$ BR($H \rightarrow ZZ^{*}$)
is estimated to be 8.3\% in the cut-based analysis and 7.9\% in the BDT analysis.

The systematic uncertainty is not taken into account in this result,
since the uncertainty is expected to be dominated by the statistical uncertainty.  
Several sources of systematic uncertainties on Higgs measurements at the CEPC is described in Ref.~\cite{higgs_mass_cepc}.
Although the study in Ref.~\cite{higgs_mass_cepc} has been performed with slightly different detector configuration and operation scenario,
the order of magnitude of these estimated systematic uncertainties $\mathcal{O}$(0.1)\%
can be also assumed for the current HZZ analysis,
that is negligible relative to the statistical uncertainty obtained from the fitting process.

\begin{table}[hbtp]
  \caption{Statistical uncertainties on the product of the \textit{ZH} cross section and the branching ratio. The bottom row shows
    the result of combined value of the six categories. }  
  \label{table:final_result}
  \centering
  \begin{tabular}{llcc}
    \hline
    \multicolumn{2}{c}{\multirow{3}{*}{Category}}  &  \multicolumn{2}{c}{ \multirow{2}{*}{$\frac{ {\Delta}({\sigma}{\cdot}BR) }{ ({\sigma}{\cdot}BR) }$ [\%]  }}  \\    
                                                                                 &        &                      &                             \\
                                                                                 &       &   cut-based    &    BDT        \\
    \hline
     \multicolumn{2}{c}{ ${\mu}{\mu}{\mathrm{H}}{\nu}{\nu}qq^{\mathrm{cut / mva}}$ } & 15  & 14  \\
     \multicolumn{2}{c}{ ${\mu}{\mu}{\mathrm{H}}qq{\nu}{\nu}^{\mathrm{cut / mva}}$ } & 48  & 42  \\
    
    \multicolumn{2}{c}{ ${\nu}{\nu}{\mathrm{H}}{\mu}{\mu}qq^{\mathrm{cut / mva}}$ } & 12  & 12  \\
    \multicolumn{2}{c}{ ${\nu}{\nu}{\mathrm{H}}qq{\mu}{\mu}^{\mathrm{cut / mva}}$ } & 23  & 20  \\    

    \multicolumn{2}{c}{ $qq{\mathrm{H}}{\nu}{\nu}{\mu}{\mu}^{\mathrm{cut / mva}}$ } & 45  & 37  \\        
    \multicolumn{2}{c}{ $qq{\mathrm{H}}{\mu}{\mu}{\nu}{\nu}^{\mathrm{cut / mva}}$ } & 52  & 44  \\    
    
    \hline
    \multicolumn{2}{l}{Combined} &  8.3 & 7.9 \\    
    \hline
  \end{tabular}
\end{table}

The signal yield ${\sigma}_{ZH}{\cdot}$BR($H \rightarrow ZZ^{*}$) combined with independently determined ${\sigma}_{ZH}$
allows the Higgs width $\mathrm{\Gamma}_{H}$ to be extracted 
as described in Sec.~\ref{section:introduction}.
Hence the precision of the Higgs width can be evaluated from the $H \rightarrow ZZ^{*}$ decay channel.
Using the following relationship
\begin{eqnarray}
  {\sigma}_{ZH}{\cdot}\mathrm{BR}(H \rightarrow ZZ^{*}) \propto g^{2}_{HZZ}{\cdot}\frac{ \mathrm{\Gamma}(H \rightarrow ZZ^{*})}{ \mathrm{\Gamma}_{H} }
  \propto \frac{ g^{4}_{HZZ} }{\mathrm{\Gamma}_{H} } \nonumber
\end{eqnarray}
the relative uncertainty of the extracted Higgs width is obtained where the relative uncertainty on square of the coupling $g^{2}_{HZZ}$ of 0.5\% 
taken from Ref.~\cite{cepc_white_paper} is assumed.
From the cut-based analysis, the relative precision of the Higgs width ${\Delta \mathrm{\Gamma}_{H}}/\mathrm{\Gamma}_{H}$ is estimated to be 8.4\%
whereas it is 7.9\% from the BDT analysis.
As mentioned in Sec.~\ref{section:introduction},  the measurement of  the $H \rightarrow WW^{*}$ decay will give another estimation on the precision of the Higgs width
in the same manner discussed above. It is shown that the precision determined from the measurement of the $H \rightarrow WW^{*}$ decay reaches 3.5\%~\cite{cepc_white_paper},
therefore, final combined precision of the Higgs width is dominated by the the $H \rightarrow WW^{*}$ measurement.
It should be mentioned that the effective field theory (EFT) is also widely accepted 
as an alternative approach to explore the Higgs couplings, where additional terms for
the interaction between Higgs and $Z$ boson in the Lagrangian collapse the simple picture above~\cite{cepc_white_paper,ilc_eft_2018}.

\section{Discussion}
\label{section:discussion}

Our estimation of the precision of the yield measurement ${\sigma}_{ZH}{\times}$BR($H \rightarrow ZZ^{*}$)
does not reach the level presented in Ref.~\cite{cepc_white_paper}.
A possible difference may exists on more sophisticated treatment of background estimation in current analysis.
Although improvement on the precision could be achieved by performing further elaborated studies to suppress backgrounds more effectively,
the improvement is also expected by considering more final states of the $H \rightarrow ZZ^{*}$ decay in the \textit{ZH} process
since only small fraction ($<3\%$) of the entire decay events has been chosen as signals and analyzed.

Broadening analysis channel of the $H \rightarrow ZZ^{*}$ decay will provide crucial qualitative improvements in studying other HZZ related topics as well.
For example, the application of EFT frameworks on the HZZ decay vertex for the study of Higgs CP properties and anomalous couplings to gauge bosons
in the presence of beyond the SM physics, has been discussed so far on the production channel ($ee \rightarrow Z^{*} \rightarrow ZH$)
for future lepton colliders~\cite{cepc_eft_study,cepc_fcc_eft_study}.
Increasing the signal statics under sufficient background rejection will allow further study on the $H \rightarrow ZZ^{*}$ vertex directly.

\section{Summary}
\label{section:summary}

The precision of the yield measurement ${\sigma}_{ZH}{\times}$BR($H \rightarrow ZZ^{*}$) at the CEPC
is evaluated using MC samples for the baseline concept running at $\sqrt{s}=240$ GeV
with an integrated luminosity of 5.6 ab$^{-1}$.
Among the various decay modes of the $H \rightarrow ZZ^{*}$,
the signal process having two muons, two jets and missing momentum
in final states has been chosen.
After the event selection, relative precision is evaluated with the likelihood fitting method on signal and background.
The final value combined from all of six categories is 8.3\% from the cut-based analysis
and 7.9\% from the BDT analysis.
The relative precision of the Higgs boson width from the $H \rightarrow ZZ^{*}$ analysis, is estimated to be 7.9\% from the BDT analysis
by combining the obtained relative uncertainty on ${\sigma}_{ZH}{\times}$BR($H \rightarrow ZZ^{*}$)
with the precision of the inclusive \textit{ZH} cross section measurement.

\begin{acknowledgements}
  The authors would like to thank the CEPC computing team for providing the simulation tools and MC samples.
  We also thank Yaquan Fang, Manqi Ruan, Gang Li, and Yuhang Tan for helpful discussions. 
\end{acknowledgements}



\begin{thebibliography}{99}
\bibitem{higgs_discovery_atlas}  ATLAS Collaboration,
  Observation of a new particle in the search for the Standard Model Higgs boson with the ATLAS detector at the LHC,
  Phys. Lett. B, {\bf 716}, 1-29 (2012).
  arXiv:1207.7214 [hep-ex]
  
\bibitem{higgs_discovery_cms}  CMS Collaboration,
  Observation of a new boson at a mass of 125 GeV with the CMS experiment at the LHC,
  Phys. Lett. B, {\bf 716}, 30-61 (2012).
  arXiv:1207.7235 [hep-ex]

\bibitem{cepc_cdr1} The CEPC Study Group, CEPC Conceptual Design Report: Volume 1 - Accelerator (2018).
  arXiv:1809.00285 [physics.acc-ph]
\bibitem{cepc_cdr2} The CEPC Study Group, CEPC Conceptual Design Report: Volume 2 - Physics \& Detector (2018).
  arXiv:1811.10545 [hep-ex]

\bibitem{cepc_white_paper} F. An {\it et al.}, Precision Higgs physics at the CEPC,
  Chinese Phys. C, {\bf 43}, 043002 (2019).
  arXiv:1810.09037 [hep-ex]

\bibitem{higgs_to_4l_atlas} ATLAS Collaboration,
  Measurement of the Higgs boson coupling properties in the $H{\rightarrow}ZZ^{*}{\rightarrow}4l$ decay channel
  at $\sqrt{s}=13$ TeV with the ATLAS detector,
  J. High Energy Phys., {\bf 03}, 095 (2018).
  arXiv:1712.02304 [hep-ex]

\bibitem{higgs_to_4l_cms} CMS Collaboration,
  Measurements of properties of the Higgs boson decaying into the four-lepton final state in pp collisions at $\sqrt{s}=13$ TeV,
  J. High Energy Phys., {\bf 11}, 047 (2017).
  arXiv:1706.09936 [hep-ex]
  
\bibitem{ilc_cdr} T. Behnke {\it et al.},
  The International Linear Collider Technical Design Report - Volume 4: Detectors (2013).
  arXiv:1306.6329 [physics.ins-det]

\bibitem{whizard} W. Kilian, T. Ohl, and J. Reuter,
  WHIZARD - simulating multi-particle processes at LHC and ILC,
  Eur. Phys. J. C, {\bf 71}, 1742 (2011).
  arXiv:0708.4233 [hep-ph]

\bibitem{mokka} P. Mora de Freitas and H. Videau,
  Detector simulation with MOKKA/GEANT4: Present and future,
  Presented at the International Workshop on Linear Colliders (LCWS 2002), 623-627 (2002),
  https://inspirehep.net/literature/609687 

\bibitem{arbor} M. Ruan {\it et al.}, Reconstruction of physics objects at the Circular Electron Positron Collider with Arbor,
  Eur. Phys. J. C, {\bf 78}, 426 (2018).
  arXiv:1806.04879 [hep-ex]

\bibitem{cepc_sample_classification} X. Mo {\it et al.},
  Physics cross sections and event generation of e$^{+}$e$^{-}$ annihiliations at the CEPC,
  Chinese Phys. C, {\bf 40}, 033001 (2016).
  arXiv:1505.01008 [hep-ex]
    
\bibitem{fastjet} M. Cacciari, G. P. Salam and G. Soyez, FastJet user manual,
  Eur. Phys. J. C, {\bf 72}, 1896 (2012).
  arXiv:1111.6097 [hep-ph]
  
\bibitem{higgs_mass_cepc} Z. Chen {\it et al.}, Cross section and Higgs mass measurement with Higgsstrahlung at the CEPC,
  Chinese Phys. C, {\bf 41}, 023003 (2017).
  arXiv:1601.05352 [hep-ex]
    
\bibitem{scikit-learn} F. Pedregosa \textit{et al.}, Scikit-learn: Machine Learning in Python,
  Journal of Machine Learning Research, {\bf 12}, 2825-2830 (2011).
  arXiv:1201.0490 [cs.LG]

\bibitem{ada_boost} Y. Freund, and R. Schapire, A Decision-Theoretic Generalization of On-Line Learning and an Application to Boosting,
  Journal of Computer and System Sciencesss, {\bf 55}, 119-139 (1997)

\bibitem{RooKeysPdf_ref} K. Cranmer, Kernel estimation in high-energy physics,
  Comput. Phys. Commun. {\bf 136} , 198-207 (2001).
  arXiv:hep-ex/0011057

\bibitem{ilc_eft_2018} T. Barklow {\it et al.}, Improved formalism for precision Higgs coupling fits,
  Phys. Rev. D, {\bf 97}, 053003 (2018).
  arXiv:1708.08912 [hep-ph]
  
\bibitem{cepc_eft_study}  I. Anderson, {\it et al.}, Constraining anomalous HVV interactions at proton and lepton colliders,
  Phys. Rev. D, {\bf 89}, 035007 (2014).
  arXiv:1309.4819 [hep-ph]

\bibitem{cepc_fcc_eft_study} N. Craig, {\it et al.},
  Beyond Higgs couplings: probing the Higgs with angular observables at future $e^{+}e^{-}$ colliders,
  J. High Energy Phys., {\bf 03}, 050 (2016).
  arXiv:1512.06877 [hep-ph]
  
\end{thebibliography}
\end{document}